# Efficient extraction of high pulse energy from partly quenched highly Er$^{3+}$-doped fiber amplifiers

**PABLO G. ROJAS HERNANDEZ,**[1,*] **MOHAMMAD BELAL,**[1] **COLIN BAKER,**[2] **SHANKAR PIDISHETY,**[1] **YUTONG FENG,**[1] **E. JOSEPH FRIEBELE,**[3] **L. BRANDON SHAW,**[2] **DANIEL RHONEHOUSE,**[2] **JASBINDER SANGHERA,**[2] AND **JOHAN NILSSON**[1]

[1]*Optoelectronics Research Centre, University of Southampton, Southampton, SO17 1BJ, United Kingdom*
[2]*Naval Research Laboratory, 4555 Overlook Ave. SW, Washington, DC 20375, USA*
[3]*KeyW Corporation, 7880 Milestone Pkwy, Hanover, MD 21076 USA*
*\*pgrh1g15@soton.ac.uk*

**Abstract:** We demonstrate efficient pulse-energy extraction from a partly-quenched erbium-doped aluminosilicate fiber amplifier. This has a high erbium-concentration, which allows for short devices with reduced nonlinear distortions, but which also results in partial quenching and thus significant unsaturable absorption, even though the fiber is still able to amplify. Although the quenching degrades the average-power efficiency, the pulse energy remains high, and our results point to an increasingly promising outcome for short pulses. Furthermore, unlike unquenched fibers, the conversion efficiency improves at low repetition rates, which we attribute to smaller relative energy loss to quenched ions at higher pulse energy. A short (2.6 m) cladding-pumped partly-quenched Er-doped-fiber with 95-dB/m 1530-nm peak absorption and saturation energy estimated to 85 µJ, reached 0.8 mJ of output energy when seeded by 0.2-µs, 23-µJ pulses. Thus, according to our results, pulses can be amplified to high energy in short highly-Er-doped fibers designed to reduce nonlinear distortions, at the expense of average-power efficiency.

## 1. Introduction

Erbium-doped fiber amplifiers (EDFAs) and lasers enable versatile and compact optical sources in the wavelength range of ~1.5 – 1.6 µm (e.g., [1-11]), and can readily be cladding-pumped with over 100 W of power from 980-nm diode lasers at low cost. Their wavelength range offers good atmospheric transmission and relative "eye-safety", highly desirable for LIDAR [2, 10, 12], remote sensing, and imaging [13]. However, the small cross-sections (absorption as well as emission) of Er$^{3+}$-ions lead to low pump absorption [14-16] and, consequently, nonlinear degradation (e.g., [2, 4, 7, 10, 17, 18]), which builds up as the signal propagates through the relatively long Er-doped fiber (EDF) (e.g., [2, 4, 10]) of, say, 10 m. Ytterbium co-doping [19-22] can then improve the pump absorption and, to some extent, the gain, in short fibers. Alternatively, a high Er$^{3+}$-concentration can also increase the pump absorption and gain, as well as the energy stored in the excited Er$^{3+}$-ions in a short fiber. However, too high Er$^{3+}$-concentrations lead to harmful quenching [14-16, 23-42], which degrades the performance. The resulting impact depends on the amplifier configuration and operating regime [6, 33, 36, 38, 41]. Notably, for high-energy pulses, the pulse duration may be shorter than, or comparable to, the quenching time scales, reported to be in the range 50 ns – 10 µs [36]. By contrast, the quenching process is often assumed to be instantaneous, but this may exaggerate the impact of the quenching on such pulses. Thus, in this regime, a finite quenching time may allow for high-energy pulse amplification at higher quenching-levels than normally considered. Indeed, refined simulations with 2-µs quenching time showed a modest pulse-energy penalty of ~10% in an EDFA core-pumped with 10-200 mW at 980 nm [43], but there was no experimental investigation. The fractional unsaturable absorption was not stated, but we evaluate it to 1.7%. This is a quite low level of quenching even compared to commercial low-concentration EDFs [42].

In this paper, we present an experimentally based investigation of the impact of quenching on amplification of high-energy pulses of 6 ns - 20 µs duration in a cladding-pumped Er-doped fiber with a high $Er^{3+}$-concentration, partly-quenched with 16.3% unsaturable absorption at 1536 nm. Our focus is not on the microscopic details of the quenching, but the effects of the quenching on amplification in this regime. Our key finding is that despite this quenching, it was possible to amplify pulses to high energy in the primarily investigated EDF. Even though the unsaturable absorption was at a level where it severely compromised the power conversion efficiency, the attainable pulse energy was comparable to that expected from an unquenched EDF. Compared to our previous conference publication [44], we now investigate shorter pulses with an improved setup.

## 2. High-Energy Pulse Amplification in Unquenched Fiber Amplifiers

In the absence of quenching, high-energy pulse amplification in rare-earth-doped fiber amplifiers, including EDFAs, has been the subject of many publications and is well understood [1-11, 17, 45-48]. In the unquenched case, amplified spontaneous emission (ASE) or spurious lasing limits the energy that can be extracted ($E_{extractable}$) in a high-energy signal pulse to a few times the intrinsic saturation energy $E_{IS}$, or say, at most 10 times [46-48] if the stored energy and the extraction efficiency are both at their practical limits. Specifically, $E_{extractable}$ is related to the gain according to [47]

$$E_{extractable} = E_{IS}\, G_{Np}^{initial} \qquad (1),$$

where $G_{Np}^{initial}$ is the initial gain in nepers when the pulse arrives. The extractable energy (which is a fraction of the energy stored in the EDFA) and thus the gain build up between pulses and generally reach their highest values when the pulse arrives. The small area of a typical core leads to low intrinsic saturation energy, so high-energy amplification in fibers requires high initial gain. However, ASE and spurious lasing limit the gain to at most ~10 Np (or ~43 dB), and thus the extractable energy according to Eq. (1) even in an unquenched amplifier. The extracted energy can be evaluated more precisely with the Frantz-Nodvik equation (FNE) [45], but the limit set by the achievable initial gain remains. Damage and nonlinearities can further limit the pulse energy as well as the peak power (e.g., [7, 17]).

## 3. Concentration Quenching

As noted above, EDFs for high-energy pulses are often highly doped to reduce the fiber length and thus the nonlinear degradation, and are therefore likely to suffer from some degree of quenching (often referred to as concentration quenching). Even if the EDF is not fully quenched and is able to reach net gain when pumped sufficiently hard, the quenching still impairs the amplification and may well limit the build-up and extraction of energy. Also quenching has been treated in many publications but there are still considerable uncertainties in the details, and considerable variations between different fibers, even if these are similar in other respects. Generally, non-radiative electric or magnetic multipolar coupling or in extreme cases even so-called direct exchange between neighboring $Er^{3+}$-ions in the metastable upper laser level ($^4I_{13/2}$) lead to quenching through energy-transfer upconversion [14-16, 23-42]. The strengths of these types of parasitic interactions depend on the distance between the $Er^{3+}$-ions in different ways, and at low concentrations, the separation between $Er^{3+}$-ions can be large enough to make quenching negligible. Tailored host glasses, e.g., co-doped with $Al_2O_3$ or $P_2O_5$ [11, 14-16, 49], as well as nanoparticle doping [40, 41, 50-53] can also counteract quenching, but it still reappears gradually at higher concentrations.

Quenching can be understood as a nonradiative lifetime shortening of the upper laser level from the unquenched value of ~10 ms. This leads to a distribution of lifetimes depending on the local environment of individual ions, but for simplicity, the ions are often grouped into two classes, isolated ions and clustered ions. Whereas the isolated ions can experience some lifetime shortening, e.g., due to so-called uniform or homogeneous upconversion (HUC) involving nonradiative energy-transfer over relatively large distances, the effects are relatively modest [14, 15, 30, 31, 34, 38, 42, 49]. It can perhaps be significant in the low-power regime, but the lifetime shortening can be fully compensated for by an increase in pump power, and the resulting pump power penalty is less significant at the powers that we used. We rarely used less than 10 W of pump power, and never used all the available pump power. In the EDF which is the focus of this paper, we did measure a lifetime of 7.7 ms, when cladding-pumping at 980 nm with pulses of 20 µs duration and 21 W peak power at 12 Hz pulse repetition frequency (PRF). This shortening from typical unquenched values of around 10 ms may well be caused by HUC of isolated ions, although we note that the decay time constant remained at ~7.7 ms also in the tail of the decay where only a small number of ions are excited and the decay constant may be expected to approach its unquenched value. Regardless, this relatively modest lifetime-shortening is essentially inconsequential. Therefore, we disregard the possible effects of HUC, including the measured modest lifetime-shortening, in this paper.

Instead, the quenching and its impact is often more strongly tied to $Er^{3+}$-clusters (e.g., pairs) [14-16, 31-35, 37, 38, 49]. In those, the nonradiative energy transfer between excited $Er^{3+}$-ions and the resulting quenching through so-called inhomogeneous upconversion is rapid, orders of magnitude faster than the fluorescence decay of unquenched ions. This makes it difficult or practically impossible to excite more than one ion in a cluster, even if the pump power is increased. Therefore, most of the clustered ions remain in the absorbing ground state, and such a cluster (or pair) forms a quenching center, or a trap. A higher $Er^{3+}$-concentration increases the concentration of clusters as well as the fraction of clustered $Er^{3+}$-ions. It is possible to measure directly the dynamics of the quenching process, by measuring the weak and short-lived fluorescence they emit. Quenching timescales have been reported to lie in the range of 50 ns – 10 µs [36], and later, by the same authors, restricted to sub-microsecond [38]. However, both the excitation and the detection used in our fluorescence measurements were inadequate for this. We note that other traps such as $OH^-$ are also possible [20], which also make it difficult to excite the $Er^{3+}$-ions. Even in cases when the relative or even absolute concentration of traps does not depend on the $Er^{3+}$-concentration, the quenching can still increase at higher $Er^{3+}$-concentrations due to increased rates of energy migration to traps [20, 26-28, 54].

Regardless of the details of such rapid quenching, it is often described in terms of a resulting unsaturable absorption [32-35, 37, 41, 42]. Unsaturable-absorption measurements directly probe the fraction of ion that are (strongly) quenched and directly quantify the unsaturable absorption, which we expect is more important for the performance of high-power EDFAs than the modest lifetime shortening characteristic of HUC (e.g., already 10% unsaturable absorption is highly significant, whereas even 50% fluorescence lifetime shortening is negligible). A signal photon emitted inside the EDF may then be lost to unsaturable absorption instead of contributing to the signal output, and an absorbed pump photon may fail to excite an $Er^{3+}$-ion. A non-zero quenching time means that the saturation characteristics become less distinct, and at sufficiently high probe power density, also the "unsaturable" absorption can saturate, if quenched ions can become excited in significant numbers. Similarly, when the fiber is used to amplify pulses of duration shorter than, or comparable to, the quenching timescale, the pulses may be able to partly saturate the normally unsaturable absorption, if the peak power and energy of the pulses are high enough. In particular, and central to this paper, an ion in the ground state that absorbs a signal photon does not have time to return to the ground state and can therefore not absorb a second signal photon in the same pulse, irrespective of whether the ion is quenched or not. Then

Eq. (1) applies to the ion collective as a whole, and the key question becomes if it is still possible to reach a high initial gain, despite the quenching. As proposed in the introduction, this may reduce the impact of the quenching and allow for high-energy pulse amplification at higher levels of unsaturable absorption than normally considered. We emphasize that the finite quenching time is essential for this hypothesis, whereas other details of the quenching process may be less important.

## 4. Experimental setup

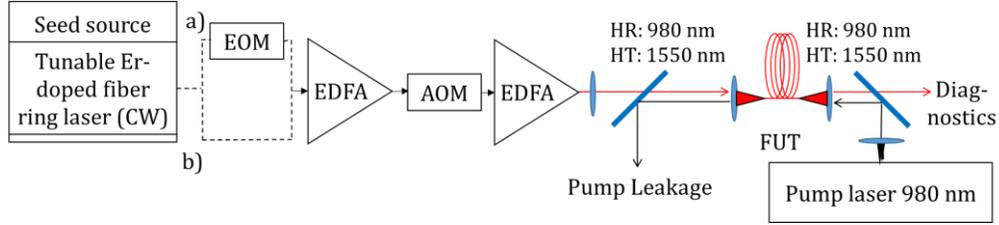

Fig. 1. Amplifier test rig for the FUT. Path (b) is used for signal pulses longer than 50 ns and up to 20 µs, defined by an acoustic optic modulator (AOM). Path (a) introduces an electro-optic modulator (EOM) to define pulses of 6 – 50 ns duration. The AOM is then set to 50 ns.

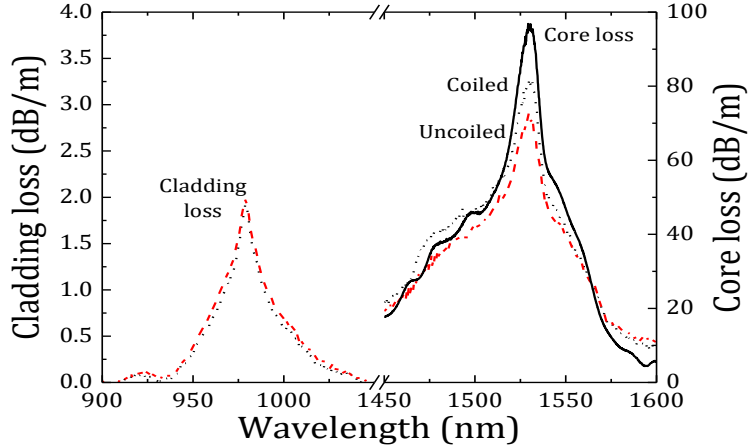

Fig. 2. White-light absorption spectra for the FUT (NRL-160415) showing core loss around 1530 nm in a short fiber (black solid curve) as well as pump waveguide (i.e., cladding) loss around 980 nm and 1530 nm for a tightly coiled (black dotted curve) and uncoiled (red dashed curve) 2.6-m-long fiber.

Our experimental layout is shown in Fig. 1. This comprises an amplified wavelength-tunable pulsed single-mode signal seed source, a pump laser, and dichroic mirrors and lenses for coupling signal and pump light into and out of the Er-doped fiber-under-test (FUT). The pump laser was a pigtailed diode laser (IPG PLD-70-974) with up to 57 W of output power at ~975 nm. The launch efficiency was ~85% into the FUT, which had both ends angle-cleaved at ~12° to suppress feedback. Counter-pumping from the signal output end was chosen for its higher efficiency in quenched EDFs [6, 31, 38].

The FUT is fabricated by Naval Research Laboratory (NRL) using MCVD and solution-doping [15,50] and is designated NRL-160415. It has a 0.13-NA, 20-µm-diameter Er-doped aluminosilicate core centered in a 125-µm-diameter circular inner cladding, which is coated by a low-index polymer. The core absorption reached 95 dB/m at the 1530-nm peak. Based on preliminary experimental data for different fiber lengths, signal wavelengths, and pulse parameters, we used a signal wavelength of 1560 nm in most amplification experiments, and an optimal length of 2.6 m. The emphasis was on high pulse energy, but this fiber length and wavelength performed well across the range of pulse parameters we used. Fig. 2 shows a white-light absorption spectrum for the core measured on a short fiber as well as for the pump waveguide (i.e., largely the inner cladding) measured on a 2.6-m long fiber. The circular symmetry of the FUT can lead to poor pump absorption. Therefore, the FUT was coiled in a way that promotes mode scrambling and thus pump absorption in one case in Fig. 2 as well as in the amplification experiments. This improved the white-light absorption in the pump waveguide of the 2.6-m piece from ~6.7 to 7.4 dB at the 1530-nm peak. These values can also be compared to the value of 95 dB/m × 2.6 m × (20 μm / 125 μm)$^2$ = 6.3 dB calculated from the core absorption and the area ratio. The values are relatively similar. The lower value calculated from the core absorption is unusual. It is possible that the core absorption is under-estimated, possibly as a result of different absorption for different modes of the core (V = 5.2 at 1560 nm). Alternatively, the probe light in the inner cladding may be disproportionately located to the center of the fiber (including the core). One possible reason for this is that the coating absorbs at 1530 nm. The absorption at the 980-nm peak becomes 5.2 dB both with and without coiling. We conclude that the pump absorption is satisfactory, without any significant mode-selective pump depletion effects. We calculated a saturation energy of 84.5 μJ from the core area and standard absorption and emission cross-sections at 1560 nm for aluminosilicate EDFs (1.69×10$^{-25}$ m$^2$ and 3.04×10$^{-25}$ m$^2$, respectively). To quantify the quenching, we measured the unsaturable absorption fraction to 16.3% with a probe at 1536 nm with up to 1 W of power, continuous-wave (CW). At longer wavelengths, the saturation power increases, and at shorter wavelengths, the available probe power was smaller. These factors hamper unsaturable-absorption measurements, and 1536 nm was the best compromise. The fiber length was 0.3 m, which gives 20 dB of small signal absorption and allowed for a maximum transmitted probe power of over 400 mW. This is well above the saturation power of unquenched ions of ~3 mW at 1536 nm, so their absorption is well saturated. For the unsaturable absorption, assuming that this is caused by ion pairs, we calculated the corresponding pair fraction to ~50% of the ions, whereas the other ~50% of the ions are isolated. The fractional unsaturable absorption with 50% paired ions becomes smaller at longer wavelengths, e.g., 13.3% at 1560 nm. Note that we use standard Er$^{3+}$:aluminosilicate cross-sections also for the quenched ions for all calculations in this paper, although deviations have been reported [39].

The seed source comprised a tunable CW Er-doped fiber ring-laser with 25 mW of output power, an optional pigtailed electro-optic modulator (EOM, Lucent 2623NA, path *a* in Fig. 1), a pigtailed acousto-optic modulator (AOM, NEOS, shortest time duration 70 ns to reach maximum transmission; 50 ns duration possible with slightly reduced transmission, extinction ratio measured to 64 dB), and two EDFAs. Optionally, the EOM was by-passed (path *b* in Fig. 1). The pulses produced by the seed source had duration of 6 ns – 20 μs and energy of 3 – 60 μJ at 1 – 40 kHz pulse repetition frequency (PRF). Path *a* was used for pulse durations of 50 ns and shorter, with the EOM running at 12 MHz and the AOM, set to 50 ns duration at the target PRF, acting as a down-sampling pulse-picker which also suppressed inter-pulse ASE. Path *b* was used for longer pulses, up to 20 μs in this paper. A dual-channel waveform generator (Tektronix AFG3252) connected to the EOM and the AOM controlled the pulse duration and PRF. The bias of the EOM was regularly adjusted to maintain an extinction ratio (ER) of ~20 dB or more.

The first EDFA (IPG EAD-5K-C) in the seed source was operated at constant current for the data we present. It yielded 900 mW of output power with 25 mW of CW-seeding at 1560 nm (path *b*), but lower output power for pulsed seeding (path *a*) with low average input power, e.g., ~5 mW with 7.2% duty cycle. Two different EDFAs were used for the second EDFA. In case of path *b*, we used an engineering prototype from SPI Lasers. In case of path *a*, we used an in-house un-packaged EDFA pumped at 1480 nm and based on 3 m of a 5-μm core EDF from Fibercore (I-15(980/125)HC). Compared to the prototype from SPI Lasers, this was better suited to the low input power to the 2$^{nd}$ EDFA that resulted with path *a*, which could be as low as 10 μW. For both path *a* and *b*, after all other parameters had been adjusted, the drive current to the second EDFA was set to produce the desired pulse energy. Whereas we focus on the useful signal energy and average power in the pulses, there was also unwanted energy between the pulses from ASE and leakage through the modulators. This includes unwanted inter-pulse energy in the seed, which limited the maximum seed pulse energy. Table 1 lists key characteristics of selected seed pulses. In the pulse energies and average powers we report in this paper, the intra-pulse contributions have been subtracted and are quoted separately. The signal launch efficiency into the core of the FUT was ~90%.

Table 1. Characteristics of selected seed pulses at 1560 nm as launched into the FUT and resulting gain and output from the FUT.

| Notes (case) | (1) | (2) | (3) | (4) | (5) | (6) |
|---|---|---|---|---|---|---|
| Seed pulse energy | 3.5 μJ | 3.5 μJ | 22.5 μJ | 4.5 μJ | 1.13 μJ | 22.5 μJ |
| Pulse duration | 6 ns | 20 μs | 1 μs | 1 μs | 1 μs | 0.2 μs |
| PRF | 2 kHz | 2 kHz | 2 kHz | 10 kHz | 40 kHz | 2 kHz |
| Energy between pulses | 0.25 μJ | 0.05 μJ | 1 μJ | 0.01 μJ | 2.5 nJ | 1.5 μJ |
| Fraction of total energy between pulses | 0.067 | 0.014 | 0.043 | 0.0022 | 0.0022 | 0.063 |
| Duty cycle | $1.2 \times 10^{-5}$ | 0.04 | 0.002 | 0.01 | 0.04 | $4 \times 10^{-4}$ |
| Seed pulse extinction ratio (7) | 61 dB | 32 dB | 44 dB | 47 dB | 41 dB | 46 dB |
| Path | *a* | *b* | *b* | *b* | *b* | *B* |
| FUT output energy | 200 μJ | 76 μJ | 770 μJ | 243 μJ | 53.7 μJ | 800 μJ |
| FUT output extinction ratio (7) | 59 dB | 21 dB | 29 dB | 31 dB | 27 dB | 36 dB |
| Resulting energy gain in FUT | 17.5 dB | 13.4 dB | 15.5 dB | 17.3 dB | 16.8 dB | 15.5 dB |
| Resulting gain for total average power in FUT | 17.7 dB | 14.8 dB | 17.5 dB | 17.3 dB | 16.8 dB | 17.5 dB |
| FUT pump power | 25 W | 25 W | 50 W | 50 W | 50 W | 50 W |

Notes
(1) Used in Fig. 4 (shortest pulse duration), and in Fig. 5.
(2) Used in Fig. 4 (longest pulse duration).
(3) Used in Fig. 6 (point with highest output energy).
(4) Used in Fig. 6 (point with highest average signal output power).
(5) Used in Fig. 6 (point with highest PRF).
(6) Used in Fig. 7 (point with highest output energy), and in Fig. 8.
(7) Ratio of average instantaneous power (energy / duration) during the pulses to that between the pulses.

Diagnostics include an optical spectrum analyzer (OSA, ANDO AQ6317B). This was used in CW measurement mode with sufficiently low measurement bandwidth to measure the time-averaged spectrum. For temporal measurements, we used a 1-GHz oscilloscope with 12 bits of vertical resolution (Agilent DSO9104H) and InGaAs photodetectors (Thorlabs DET10C, 35 MHz, detector area 0.8 mm$^2$, bias voltage 5 V, and EOT ET-3500, ~15 GHz, detector area $8 \times 10^{-4}$ mm$^2$). Since the leakage through the modulators in the seed occurs largely at the seed wavelength, we determined the pulse energy from the average power and oscilloscope traces and checked there were no inconsistencies in the optical spectrum (e.g., excessive ASE). See Appendix for details. Average optical powers were measured with thermal and semiconductor power meters.

## 5. Results and Discussions

Fig. 3 shows the gain for a CW seed with 1 mW of launched power at 1560 nm. This reaches 34 dB = 7.8 Np. The pump leakage was ~30% (–5.2 dB) at high gain. Thus, the pump absorption is similar to the white-light absorption at the 980-nm peak, despite the significant excitation of $Er^{3+}$-ions, and any non-ideal spectral overlap between the pump and the absorption. A possible explanation is that the spatial overlap with the $Er^{3+}$-ions is larger for the pump light than for the white light used in Fig. 2. The gain slope in Fig. 3 drops at high pump power due to gain saturation, as the signal output power exceeds 2 W. Lower seed power reduces the saturation, but 1 mW is reasonably representative of the seed's parasitic inter-pulse power, so the measured gain is an indication of what initial gain may be achievable (i.e., $G_{Np}^{initial}$ in Eq. (1)). Note however that during high-energy pulse amplification, the instantaneous gain varies continuously in time, and the precise initial gain was not measured in the pulsed experiments. The differential conversion efficiency in the saturated regime becomes 12% with respect to absorbed pump power for the data in Fig. 3. Simulations gave a differential conversion efficiency of 18% with 50% of paired ions. The quenching of pairs was assumed to be instantaneous in these simulations, since it is expected to be much faster than $Er^{3+}$ absorption and stimulated-emission rates for the CW data in Fig. 3, which correspond to a typical time scale of, e.g., 0.1 ms. Without quenching, the simulated differential conversion efficiency was at the quantum limit of 63%, which is ~7 dB higher than the experimental value. This underlines the strong detrimental impact of quenching in the CW regime. Nevertheless, the high gain that we reach may allow for high energies, according to Eq. (1).

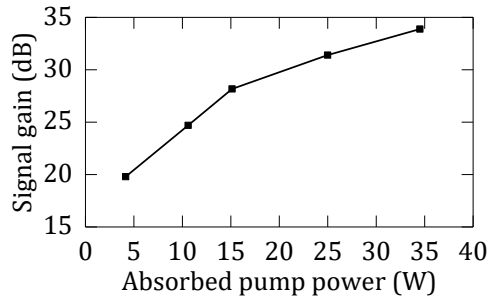

Fig. 3. Gain vs. absorbed pump power for 1 mW of CW input signal at 1560 nm.

Fig. 4 shows the output signal pulse and inter-pulse energy vs. pulse duration for 2-kHz PRF, 3.5-μJ seed energy (average seed power 7 mW), and 25 W of launched pump power (17 W absorbed). The pulse duration varies from 6 ns to 20 μs, and the seed's peak power from ~580 W down to ~0.17 W. The EOM (path *a*) is used only for durations of 50 ns and less. The total output power as measured with a thermal power meter is also shown, in terms of the total energy during the pulse period, i.e., the average power divided by the PRF. For the longest pulses (20 μs), the output energy reaches 68 μJ, so the energy gain becomes 13 dB. The total average output power becomes 190 mW. The ER becomes 21 dB in terms of the power during the pulses relative to the average power between pulses. The stimulated-emission rates induced by the input and output signal pulses become ~1290 $s^{-1}$ = (770 μs)$^{-1}$ and ~25,100 $s^{-1}$ = (40 μs)$^{-1}$, respectively, if we assume that the pulses are rectangular. All these values then increase for shorter pulses. Thus, for 6-ns pulses, the energy gain increases by 5 dB from that of the 20-μs pulses and reaches 18 dB. The ER becomes 59 dB, and the total average output power becomes 450 mW. The output pulse energy reaches 0.2 mJ energy (~2.4 times the saturation energy). For comparison, the FNE yields 0.27 mJ of output pulse energy with standard cross-sections for an unquenched EDF and 28 dB of initial gain, obtained from Fig. 3 for 17 W of absorbed pump power. This is only 1.5 dB higher than the experimental result for 6-ns pulses. The stimulated-emission rates induced by the input and output signal pulses become ~4.3×10$^6$ $s^{-1}$ = (230 ns)$^{-1}$ and ~$250\times10^6$ s$^{-1}$ = (4.1 ns)$^{-1}$, respectively, if we assume that the pulses are rectangular. This is comparable to, or faster than, reported quenching rates [36, 38].

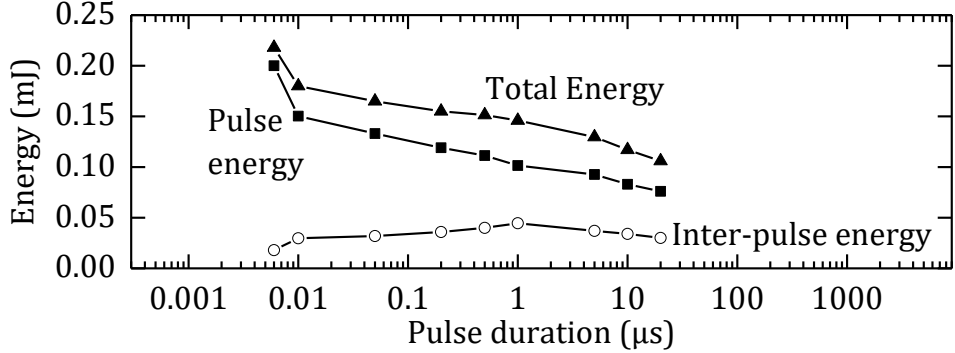

Fig. 4: Output pulse, inter-pulse, and total energy vs. pulse duration for seed energy of ~3.5 µJ, PRF of 2 kHz, and launched pump power of 25 W (17 W absorbed).

Fig. 5 (a) shows the temporal profile of the 6-ns pulse that produced the highest energy of 0.2 mJ in Fig. 4. We reach 40 kW of peak power. Fig. 5 (b) shows the average-power spectrum. With 40 kW of peak power, the nonlinear effect of four-wave mixing generates sidebands (e.g., [7, 17, 18]) containing ~44% of the energy. Sideband energy is included in all reported energies, but is negligible in most cases.

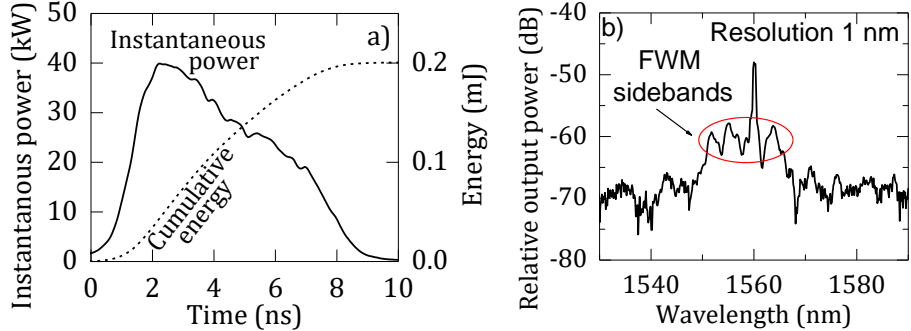

Fig. 5. (a) Instantaneous signal output power and cumulative pulse energy from the beginning of the pulse and (b) optical spectrum for 6-ns pulse duration at 17 W of absorbed pump power. Case (1) in Table 1.

When we take into account the uncertainties resulting from possible differences between quenched and unquenched erbium cross-sections [39, 43] and in other experimental parameters, as well as the spectral broadening, Fig. 4 suggests that for sufficiently short pulses, it is possible to recover the energy gain of unquenched fibers, as described by the FNE, also in our quenched fiber. This is a key result of our experiments. As outlined briefly above, we propose the following explanation: The dynamics of Er$^{3+}$-ions may be slow compared to the pulse duration. Then, all transitions during the pulse are negligible, compared to those induced by the high-energy signal pulse. This applies to unquenched as well as quenched ions, insofar as the quenching dynamics are slower than the pulse duration. Thus, even if the absorption of quenched ions (in the ground state) is effectively unsaturable in the CW regime, it behaves as a saturable absorber with a short pulse, just like unquenched ions in the ground state do. If other differences in the spectroscopy of quenched and unquenched ions are small, it no longer matters to the pulse if the ions in the absorbing ground state are quenched. Partly quenched and unquenched fibers should then yield similar output pulse energy for the same initial gain. By contrast, for longer pulses, a quenched ion that absorbs a signal photon has time to lose its energy through parasitic nonradiative processes during the pulse and return to the ground state. It can then absorb another signal photon, leading to lower output energy. Thus, the measured output energy recovers its unquenched value for shorter pulses in Fig. 4, but decreases significantly for longer pulses. Such dependence is not expected for unquenched fibers. We hypothesize that it also helps to maintain the ER for shorter pulses, but not for longer pulses (cf. Table 1), although further investigations would be needed to confirm that.

Although Fig. 4 shows higher output energy for shorter pulses, this was with pulse seed energy limited to ~3.5 µJ and with non-negligible inter-pulse seeding. Therefore, for the seeding used in Fig. 4, higher pump power than 17 W (absorbed) rapidly increased the energy between the pulses rather than in the pulses. Higher seed energy may lead to higher output energy, although this was beyond the capability of our seed source for the 6-ns case in Fig. 4 and 5. With longer pulses, however, the seed source can reach higher pulse energy with acceptable inter-pulse energy and power, thanks to the lower gain of the amplifiers. Thus, the FUT can be pumped with higher power and reach higher initial gain, which opens up for higher output pulse energy. Fig. 6 (a) depicts the average output power and pulse energy vs. PRF for a constant average seed power of 45 mW in 1-µs seed pulses at three wavelengths, 1555 nm, 1560 nm and 1565 nm, with 50 W of launched pump power (35 W absorbed). The PRF varied from 1 to 40 kHz, so the corresponding seed energy varied from 45 µJ down to 1.1 µJ. The results across the three wavelengths show a ~6 dB increase in average output power for an increase in PRF from 1 kHz to 10 kHz. This is expected, because of decreasing energy saturation and inter-pulse ASE for higher PRF. For an unquenched system, one expects this trend to continue with a further increase in average output power also for PRF above 10 kHz, asymptotically towards a maximum for high PRF. Instead, Fig. 6 (a) shows a small drop in power. We propose this is another, more subtle, result of quenching. Even if a short pulse saturates the absorption of the quenched ions by exciting them for the duration of the pulse, the energy in the excited quenched ions is then rapidly dissipated through the quenching process. The saturation implies that the absorbed energy increases sub-linearly with pulse energy, so lower-energy pulses at higher PRF increase the fraction of the pulse energy, and thus the average power, deposited into the quenched ions. This reduces the average output power for higher PRF.

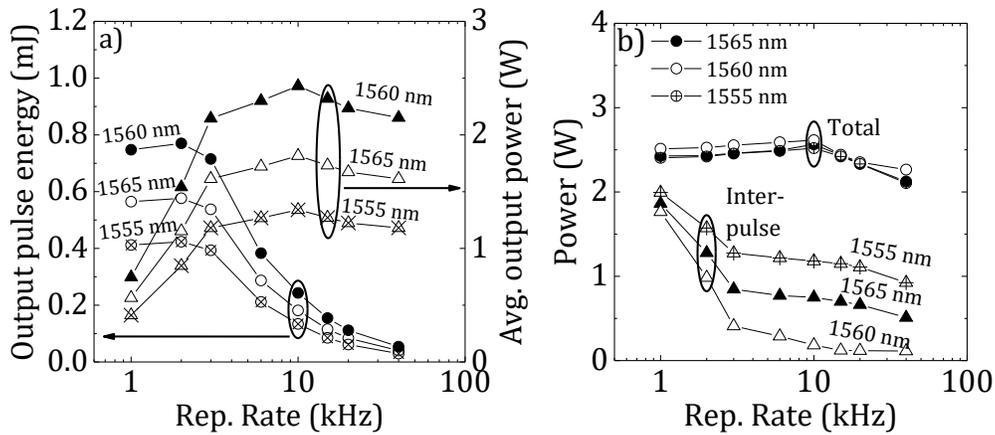

Fig. 6. (a) Output pulse energy and average signal power *vs*. PRF at constant average signal input power of 45 mW in 1-µs pulses at three wavelengths, 1555 nm, 1560 nm and 1565 nm with 50 W of launched pump power (35 W absorbed). (b) Total output power (in pulses + between pulses) and inter-pulse power.

The decrease in power for higher PRF is confirmed in Fig. 6 (b), which shows the total output power as measured with a thermal power meter. It also shows the average inter-pulse power. In conventional fashion, the inter-pulse power is high at low PRF and lower for higher PRF. The inter-pulse power at 1555 nm and 1565 nm is significantly higher than at 1560 nm, especially at high PRF. This leads to an unusually strong wavelength dependence in Fig. 6 (a). For PRF lower than 3 kHz, the rapid increase in inter-pulse power hampers further pulse-energy growth, and the inter-pulse power reaches over 45% of the average signal power already at 2 kHz. Although ASE is generally bi-directional, the FUT is seeded in the forward direction by inter-pulse power from the seed laser (ASE as well as leaked signal). Compared to the forward inter-pulse energy, we expect backward-propagating inter-pulse power to be negligible.

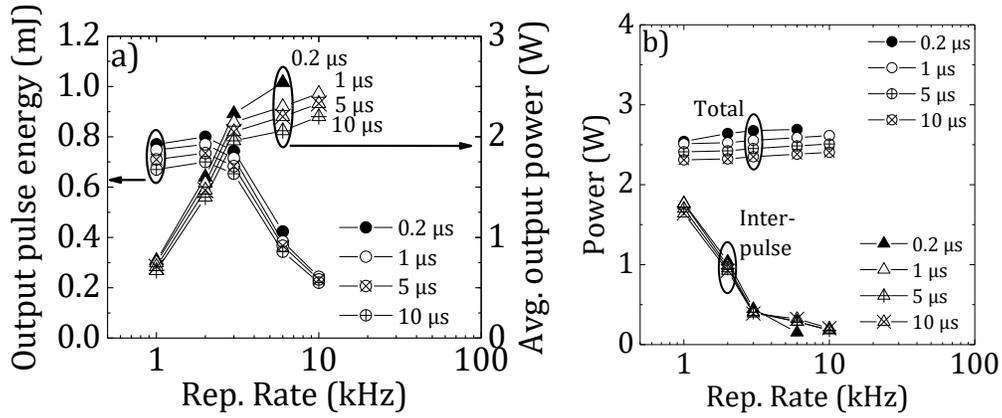

Fig. 7 (a) Output pulse energy and average signal power vs. PRF at constant average signal input power of 45 mW at 1560 nm for different pulse durations with 50 W of launched pump power (35 W absorbed). (b) Total output power (in pulses + between pulses) and inter-pulse power.

We next consider the effect of pulse duration in this regime of high-energy seeding and extraction. For this, the parameters are the same as in Fig. 6, except that we seed at 1560 nm and vary the pulse duration from 0.2 μs to 10 μs with PRF in the range 1 to 10 kHz. The duty cycle varies between $2\times10^{-4}$ (0.2 μs, 1 kHz) and 0.1 (10 μs, 10 kHz). The seed energy varies between 4.5 μJ at 10 kHz and 45 μJ at 1 kHz. The average power during the seed pulses varies between 0.45 W and 225 W (this is similar to the peak power of the seed pulses, but at high seed energy, saturation-induced pulse shortening can be significant already in the seed). The results are shown in Fig. 7. We see that the trend with respect to PRF is the same as in Fig. 6. Furthermore, the trend with respect to pulse duration is the same as in Fig. 4, i.e., the shortest pulse of 0.2 μs leads to the highest energy, which is reached at 2 kHz rather than at 1 kHz, although the difference in output energy may be too small to be significant. Thus, with 2-kHz PRF seeding in 0.2-μs pulses (so with 22.5 μJ energy and 110 W average power and ~$0.84\times10^6$ $s^{-1} = (1.2\ \mu s)^{-1}$ stimulated-emission rate during the seed pulse), the maximum pulse energy from the FUT reaches as high as 0.8 mJ, despite a conversion of no more than 4.6% of absorbed pump power. Fig. 8 (a) shows the corresponding instantaneous power and cumulative pulse energy, and Fig. 8 (b) shows the spectrum. The FWHM duration becomes 110 ns, and the actual peak power 6.2 kW, leading to a stimulated-emission rate of ~$47\times10^6\ s^{-1} = (21\ ns)^{-1}$. The energy between pulses becomes 53 μJ and the ER 36 dB. The spectral purity is relatively good, with 89% of the power at the signal wavelength. The energy gain reaches 15.5 dB. The energy of 0.8 mJ is 9.5 times the estimated saturation energy, and is 1.5 dB higher than the 0.57 mJ calculated with the FNE, with initial gain of 34 dB (estimated from Fig. 3) and seed energy of 22.5 μJ. Thus, as for the 6-ns pulse in Fig. 5, the agreement with the FNE is fair, but in contrast to the 6-ns pulse, the energy is now higher than that predicted by the FNE. Possible contributions to this difference include energy measurement errors, errors in the estimate for the initial gain used in the FNE, the increased ability of a higher-energy seed pulse to extract energy also in the edges of the core where the signal intensity is relatively low as well as from any $Er^{3+}$-ions with atypically small cross-sections, and the increased ability of a higher-power pump to excite paired $Er^{3+}$-ions, which may have smaller cross-sections [39, 43] and thus store more energy for a given contribution to the gain. We also note that measurements with the same seed energy and PRF on a low-quenched EDF (unsaturable absorption 4.5% at 1536 nm) agreed well with the FNE. This EDF was also fabricated by NRL but with nanoparticle-doping [53] and a lower Er-concentration (peak core absorption 33 dB/m) to avoid quenching.

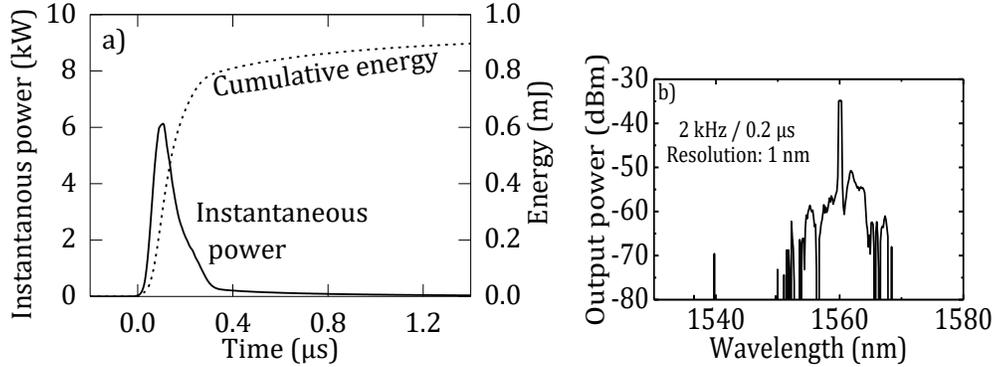

Fig. 8. (a) Instantaneous power and cumulative energy at highest energy and (b) output spectrum. Case (6) in Table 1.

Even if the effect of the quenching on the extraction process is small and the pulse energy is well described by the FNE, it is possible that the erbium excitation reached in the pumping process, and thus the initial gain, is more strongly affected by quenching. Based on the CW signal gain of 34 dB at 1560 nm for 35 W of absorbed pump power, we roughly estimate that 55% of the $Er^{3+}$-ions are excited when a pulse arrives. This is a relatively low percentage, which may partly be a result of quenching-induced degradation of the pumping process. Under the assumption of equal spectroscopic parameters for isolated and paired ions, we estimate that 40% of the paired ions and 70% of the isolated ions are excited, as averaged over the fiber, when a pulse arrives. Still, even if the pumping is degraded, the 2.6-m long partly-quenched EDF outperforms unquenched silica EDFs (e.g., [1, 4]) in generation of energy per unit core area and length (0.98 $\mu J/\mu m^2/m$) and gain per unit length (>5.9 dB/m) at 1560 nm. Although the conversion efficiency decreases, our results show that readily available pump power is more than enough to make the pulse energy limited by radiative losses to parasitic inter-pulse emission (including ASE) rather than by nonradiative quenching losses.

Although we have deliberately selected an EDF that is partly quenched for our studies, similar $Er^{3+}$-concentrations are possible with less quenching, as demonstrated with a $P_2O_5:Al_2O_3:SiO_2$ glass matrix [11] (we estimate the fraction of unsaturable absorption to be half of that in our fiber). This allowed for pulse amplification with relatively high efficiency (8.4%) in a 1.7-m long EDF with a 35-$\mu m$ core to 84 kW in 2-ns pulses [11] (energy extraction ~0.1 $\mu J/\mu m^2/m$). Thus, it is clear that the relation between $Er^{3+}$-concentration, quenching, and unsaturable absorption is not unique. Investigations of a second partly-quenched EDF showed that the relation between the (fractional) unsaturable absorption and the impact on high-energy pulse amplification is not unique, either. Although it had a lower fraction of unsaturable absorption (12% at 1536 nm) than the first partly-quenched EDF presented in this paper, we could not extract high pulse energy. It is possible that it suffers from exceedingly fast quenching by direct exchange [23, 24]. This was an experimental nanoparticle-doped aluminosilicate EDF [53]. Nanoparticle-doping can reduce the quenching, but we selected this EDF for its unusually high quenching, and it is possible that also the quenching characteristics differ from those in

EDFs fabricated in other ways. A hypothesis is that when the fabrication of a NP-doped fiber preform is unsuccessful, it leads to pairs or clusters of $Er^{3+}$-ions with very small separation and thus to fast quenching. We also mention that for paired ions, it is possible to excite half of the ions even if the quenching process occurs instantaneously (when both ions become excited). It is therefore possible to reach gain also in such paired ions for wavelengths longer than the zero-phonon wavelength of 1530 nm. At 1560 nm, this requires a 0.98-μm pump intensity of 0.13 mW/μm$^2$, to excite 36% of the paired ions, under the assumption of standard cross-sections and unquenched lifetime, and instantaneous quenching. This intensity translates to 1.5 W of pump power in our fiber. Higher pump power as well as a non-zero quenching time are expected to increase the gain from paired ions, whereas larger clusters would be expected to exhibit less gain (if any), under the assumption that only one ion per cluster can be excited. Detailed spectroscopic investigations beyond the scope of this work are needed to evaluate the impact of such factors.

## 6. Conclusions

In summary, our results demonstrate that short pulses can efficiently extract high energy from a partly-quenched high-concentration erbium-doped fiber amplifier, even though the fiber exhibits significant unsaturable absorption in the CW regime. We reached up to 0.8 mJ of output energy from a short (2.6 m) EDFA, which is 9.5 times the estimated saturation energy. Although such results have not been reported before as far as we are aware, the high energy extraction in a short pulse can be readily understood in terms of the dynamics of the quenching process. Thus, we attribute the high energy achievable in this regime to the rapid extraction of stored energy, on time scales faster than the quenching dynamics. Thereby, the short high-energy pulses can saturate the absorption of ions which are unsaturable in the CW regime. This implies that insofar as it is possible to reach a high small-signal gain, it is possible to generate pulse energies of several times the saturation energy. We reached output energies within 1.5 dB of those predicted by the Frantz-Nodvik equation for the unquenched case. Furthermore, in some pulse regimes, the average-power conversion efficiency increased in ways that would not be expected for an unquenched EDF. On the other hand, the impact of the quenching on the average-power conversion efficiency was large, and led to a 7-dB degradation in the CW regime. Furthermore, the amplification of high-energy pulses was severely compromised in another EDF with partial quenching, and we hypothesize that the quenching timescale may be much faster in that fiber. Further studies of the details of different quenching and parasitic processes (which may involve higher-lying energy levels [42]) in different regimes are needed to better understand the impact on high-energy amplification, and how it may depend on composition and fabrication details. Particularly interesting is to what extent the positive results of high-energy pulse amplification in partly-quenched EDFs reported here carry over to hosts and fabrication approaches known for low quenching, at even higher Er-concentration where significant quenching reappears also in such hosts.

**Appendix A: Determination of Pulse Energy From Oscilloscope Traces**

We measured temporal traces with a detector and an oscilloscope to determine the shape, duration, and energy of our pulses. The shape and duration are provided directly by such measurements, for which a dynamic range of 20 dB is typically adequate and easily obtainable. We measured pulse shapes with the 15-GHz detector (EOT ET-3500) connected directly to the oscilloscope with 50-Ω termination and 1-GHz bandwidth.

If the average power is measured separately, it is in principle straightforward to calibrate the oscilloscope trace in terms of instantaneous power and then from that also determine the peak power and, through integration, the pulse energy. However, although conceptually simple, this is often challenging, especially at low PRF, where the long interval between pulses can lead to significant inter-pulse energy even at low inter-pulse power in the form of ASE and / or leaked signal. Inter-pulse power is a common issue with high-energy fiber sources, because of the high (initial) gain they reach. This is a problem for the assessment of pulse energy as the average power divided by the PRF, which fails to exclude the energy between pulses, as well as through the integration of the pulse trace, because the low duty cycle means that the inter-pulse signal needs to be measured accurately even when it is a very small fraction of the peak power. Consequently, given the level of sensitivity and accuracy required to determine the pulse-energy at low duty cycle, it is sometimes even claimed that the ASE power cannot be detected by a standard photodetector [9]. To overcome this problem, an AOM can be used as a time-gate [8] to separately measure the average power during and between pulses. Alternatively, instead of numerical integration of the digitized oscilloscope trace in a computer, one can integrate the photodetector signal in an analog electronic circuit [9]. This has been found to work well for 100-ns pulses at 10-kHz PRF with a bespoke circuit, although at a relatively small dynamic range (on-off ratio) of ~30 dB [9].

In our case, we determined the pulse peak power to be as high as 60 dB above the inter-pulse power, for the 0.2-mJ, 6-ns pulse in Fig. 5(a) (case (1) in Table 1). This is much more than the dynamic range of oscilloscopes. In order to determine the pulse energy with such high on-off ratios, we reduced the bandwidth with a low-pass circuit comprising a 4.7 nF capacitor in parallel with a 150-$\Omega$ resistance, 50 $\Omega$ of which was in the oscilloscope's input port. This results in a bandwidth calculated to ~226 kHz, which agrees well with the filter's measured time constant of ~682 ns. Note that for a pulse significantly shorter than the filter's time constant, the peak signal voltage is given by the charge generated by the pulse divided by the capacitance, and is therefore proportional to the pulse energy. Thus, the 0.2-mJ, 6-ns pulse in Fig. 5(a), would generate a voltage of 0.2 mC / 4.7 nF = 42.5 kV, if we assume a detector responsivity of 1 A/W = 1 C/J. This voltage is excessive. We therefore attenuated the light incident on the detector to yield a peak voltage of 0.3 V over the oscilloscope's 50 $\Omega$ (and another 0.6 V over the 100-$\Omega$ resistor in series), so by ~46.7 dB for the pulse in Fig. 5(a). (Since the voltage is proportional to the optical power or energy, we use a multiplier of 10 in the dB scale rather than 20, as normally used for voltages.) We attenuated the light reaching the detector by passing it through a multimode patchcord, which collected a small fraction of the output signal. For this, the signal was first passed through a diffuser, so that the collected light is representative of the whole beam. The use of a patchcord also shields the detector from ambient light.

We next consider the inter-pulse power, which we determined to ~40 mW. Given the attenuation of ~46.7 dB, this results in a photocurrent of ~847 nA and voltage of 42.4 µV over the oscilloscope's 50 $\Omega$. Although the low-pass filter significantly reduces the signal's dynamic range, it is still quite high, i.e., 0.3 V / 42.4 µV or ~38.5 dB. It would be possible to reduce this further by increasing the filter's time constant. However, this increases the time it takes for the signal to decay from its peak value. In our case, it takes around 38.5/4.343 = 8.87 times the time constant of 682 ns (i.e., 6.05 µs) for the signal to decay from the peak level to the inter-pulse level for a first-order filter. It is not possible to measure the instantaneous power during this time. A time constant which is, say, ten times longer, would increase the black-out period to approximately 6.82 µs × 28.5/4.343 = 44.8 µs, which we view as excessive. By contrast, we consider the uncertainty in the pulse energy created by the black-out period of our 226-kHz low-pass filter to be acceptable even at 40 kHz (where the inter-pulse energy is considerably smaller).

The signal's dynamic range after the low-pass filter is determined entirely by the filter bandwidth. It is also necessary to keep the signal within the linear range of the measurement system. The lowest measurable signal, i.e., the lower end-point of the linear range, can be limited by noise, quantization errors, and offset errors. With the measurement range set to 0 – 0.4 V (50 mV per division), the RMS noise of our oscilloscope in 1 GHz is stated as 550 µV. Although this exceeds the inter-pulse signal, our objective is to evaluate the inter-pulse energy, i.e., the integral of the signal. Integration over the 0.5-ms inter-pulse span for Fig. 5(a) reduces the RMS noise to ~0.78 µV (if we assume white noise). This is well below the signal level. As it comes to quantization with 12 bits and 0.4 V full range, the least significant bit (LSB) corresponds to 100 µV. This exceeds the inter-pulse signal level. However, the noise is of the order of the LSB voltage or more. Under these conditions, averaging reduces the quantization error of the averaged signal. We evaluated it to be less than 1 µV, which is negligible.

As it comes to the detector, the noise equivalent power is negligible (<< 1 nW) already at the filter bandwidth, and is reduced even further by the integration. Furthermore, the dark-current is specified to 1 – 25 nA. This is small or negligible compared to the average inter-pulse photo-current, and furthermore, we measured and subtracted dark traces in all cases.

With these steps, we believe we may be able to measure the average inter-pulse signal at the µV-level, i.e., within 2.4% the average of 42.4 µV, and at least within 10%.

The linear range is also limited upwards, e.g., by saturation. To explicitly investigate linearity, we attenuated the optical signal and investigated the effect on the signal voltage in the pulse as well as between pulses, under representative measurement conditions, and starting with a peak voltage of ~0.3 V. We found that the deviations from linearity were smaller than 1 dB for 30 dB of dynamic range under representative measurement conditions with 1 MHz filter bandwidth. This may lead to an error in the inter-pulse energy of ~20%, which may seem excessive. However, insofar as the inter-pulse energy is relatively small, this does not lead to excessive errors in the pulse energy. For example, Fig. 4 shows that the inter-pulse energy is ~20 µJ, i.e., ~10% of the total energy. Even if the deviations from linearity causes a 20% error in this energy, this only corresponds to 4 µJ or 2% of the pulse energy of 0.2 mJ. This error is acceptable. We note that the impact of a relative error in inter-pulse energy becomes more significant at higher levels of inter-pulse energy. Although the inter-pulse energy exceeded the pulse energy in some cases, this was for longer pulse durations, for which deviations from linearity are expected to be smaller.

Finally, we mention also that there are relatively straightforward ways to improve the accuracy. For example, oscilloscopes can readily average over several traces, and we did on occasion use modest such averaging. Furthermore, pulses can be measured with different oscilloscope gain settings, as well as with different filter bandwidths. The power onto the detector can be varied, too. During the experiments, we regularly changed these parameters to better understand the pulse characteristics and check for anomalies. However, the traces from which we calculated the pulse energy were generally measured with ~0.3-V peak voltage and ~226-kHz filter bandwidth. Overall, we estimate errors in reported pulse energies to roughly 10% or less.

All data supporting this study are available from the University of Southampton at https://doi.org/10.5258/SOTON/D0853.

**Acknowledgment**
We acknowledge John Marciante, University of Rochester, for helpful discussions on detector linearity and pulse measurements, and Michalis Zervas, University of Southampton, for suggestions on the manuscript.

**Funding**
University Research program of Northrop Grumman Mission Systems and National Council for Science and Technology (CONACyT), Mexico (578927/409382).

**Disclosures**
The authors declare no conflicts of interest.